\documentclass[aps,twocolumn,prl,showpacs]{revtex4}
\usepackage{graphicx,amssymb,amsmath}
\usepackage{graphicx}
\usepackage{dcolumn}
\usepackage{bm}

\newcommand{\Sxx}{$S_{xx}$}

\begin{document}

\title{Thermoelectric response of fractional quantized Hall and re-entrant insulating states in the N=1 Landau level}

\author{W.E. Chickering$^1$, J.P. Eisenstein$^1$, L.N. Pfeiffer$^2$, and K.W. West$^2$}

\affiliation{$^1$Condensed Matter Physics, California Institute of Technology, Pasadena, CA 91125
\\
$^2$Department of Electrical Engineering, Princeton University, Princeton, NJ 08544}

\date{\today}

\begin{abstract} Detailed measurements of the longitudinal thermopower of two-dimensional electrons in the first excited Landau level are reported.  Clear signatures of numerous fractional quantized Hall states, including those at $\nu = 5/2$ and 7/3, are observed in the magnetic field and temperature dependence of the thermopower.  An abrupt collapse of the thermopower is observed below about $T= 40$ mK at those filling factors where re-entrant insulating electronic states have been observed in conventional resistive transport studies.  The thermopower observed at $\nu = 5/2$ is discussed in the context of recent theories which incorporate non-abelian quasiparticle exchange statistics.
\end{abstract}

\maketitle

The thermoelectric power of low dimensional conductors provides a view of such systems that is distinct from that offered by conventional charge transport.  For example, it has been shown that in the disorder-free limit the longitudinal thermopower \Sxx\ of a two-dimensional electron system (2DES) at high magnetic field is given by $S_{xx} = -S/(|e|N_e)$, where $S$ is the system entropy, $N_e$ the number of electrons in it, and $e$ the electron charge \cite{obraz,cooper,shastry}.  The potential for access to such a basic equilibrium thermodynamic variable by means of a non-equilibrium thermal transport measurement is particularly enticing in strongly correlated low dimensional systems where unusual electronic degrees of freedom often exist.

The strongly correlated electron system of interest here is a 2DES in which a perpendicular magnetic field $B$ has positioned the Fermi level in the first excited ($N=1$) orbital Landau level.  It is well-known that such a system displays a remarkably diverse array of unusual collective electronic phases. Conventional transport measurements on this system have established the existence of several fractional quantized Hall effect (FQHE) states, including the even-denominator state at Landau level filling fraction $\nu = nh/eB = 5/2$ (with $n$ the electron density and $h$ Planck's constant) and the odd-denominator states at $\nu = 7/3$ and 8/3 \cite{willett,jpe90,pan08}.  In the highest mobility samples intriguing insulating phases with $integer$ Hall quantization have also been observed lying in between various FQHE liquid states \cite{riqhe}.  The situation is further enriched by the application of an in-plane magnetic field component \cite{jpe88,jpe90,csathy05,gervais,xia10,xia11}.  The in-plane field destroys the $\nu = 5/2$ FQHE and, initially at least, replaces it with an anisotropic compressible phase similar to those found in the $N \geq 2$ Landau levels \cite{lilly1,du,pan,lilly2}.

In this paper we report detailed measurements of \Sxx\ in the $N = 1$ Landau level in a 2DES of extremely high quality.  Our measurements extend down to about $T = 20$ mK, where the thermopower is strongly dominated by electron diffusion rather than phonon drag effects.   In this regime clear signatures of almost all of the known FQHE states in the $N = 1$ Landau level are observed in \Sxx.  Below about $T = 40$ mK the above-mentioned insulating phases appear and abruptly quench the thermopower.  In the vicinity of the $\nu = 5/2$ FQHE we find \Sxx\ to be in rough quantitative agreement with recent theories which incorporate the enhanced entropy expected from non-abelian quasiparticle exchange statistics. 

The sample used in this experiment is a modulation-doped GaAs/AlGaAs heterostructure grown by molecular beam epitaxy on a $\langle 001 \rangle$-oriented GaAs substrate.  The 2D electron gas resides in a 30 nm quantum well sandwiched between thick layers of Al$_{0.24}$Ga$_{0.76}$As and buried 210 nm below the sample top surface. After low temperature illumination with red light, the 2DES has a density of $n=2.9 \times 10^{11}$ cm$^{-2}$ and a low temperature mobility of about $\mu = 3 \times 10^7$ cm$^2$/Vs.  A rectangular bar, 6 mm wide by 12 mm long, is cleaved from the parent wafer and then thinned, from the substrate side, to about 130 $\mu$m thickness.  Chemical etching the top surface confines the 2DES to two independent 3 mm square mesas positioned along the bar. Six AuNiGe ohmic contacts are placed at the corners and two side midpoints of each mesa.  A thin film heater, of serpentine shape, is deposited at one end of the bar and covers most of the bar's width.  Narrow evaporated Ti/Au contact lines run from each of the ohmic contacts to the 2DES mesas and from the heater to a patch bay at the opposite end of the bar.  This end of the bar is In-soldered to the cold finger of a dilution refrigerator and serves to define thermal ground. The inset to Fig. 1 illustrates the sample layout.

As reported previously, the 2DES functions both as the thermoelectric material of interest and as an $in$-$situ$ thermometer \cite{chickering}.  At magnetic fields close to strong integer quantized Hall states, the longitudinal resistivity $\rho_{xx}$ of the 2DES provides a sensitive thermometer which is used to calibrate the phonon-dominated thermal conductances $K_{1,2}$ of the bar between each 2D mesa and thermal ground.  We find that these conductances scale with temperature as $\sim T^{2.6}$ for $T \gtrsim 40$ mK and depend as expected on the distances between the 2DES mesas and the thermal ground \cite{geometry}.  The phonon mean free path extracted from these conductance measurements is weakly temperature dependent but otherwise broadly consistent with nearly boundary scattering-limited phonon transport \cite{wybourne}.  Although the conductance measurements are made in the presence of a large magnetic field, we find no evidence for any systematic dependence upon the field.  The conductances $K_{1,2}$ allow us to determine the temperature profile along the bar when a heat flux is applied.

In contrast to our previously reported quasi-dc thermopower measurements, which were hampered by long thermal relaxation times ($\sim 10-100$ sec) of the sample bar, we here report precision low frequency ac results of much higher quality and at considerably lower temperatures.  This is possible owing to the far shorter thermal relaxation time ($< 1$ msec) of our present device \cite{timeconstant}.  By applying an ac current (at typically $f = 13$ Hz) to the serpentine heater, a heat flux and temperature gradient along the bar are created at frequency $2f$. The induced longitudinal thermoelectric voltages in the 2DES are lock-in detected at $2f$.  No frequency dependence or anomalous phase shifts of the thermoelectric voltages were observed at these low frequencies. Care was taken to ensure that the temperature drop $\Delta T$ across either 2DES mesa was always less than 10$\%$ of the mean temperature of the 2DES.  Dividing the measured thermoelectric voltage by $\Delta T$ yields the longitudinal thermopower \Sxx. We concentrate here on the thermopower data obtained from the 2DES mesa closest to the thermal ground; except in the few instances discussed below, essentially identical results were obtained from the second 2DES mesa.

Before turning to the thermopower in the $N=1$ LL, we discuss our findings at filling fraction $\nu = 3/2$.  Here the upper spin branch of the $N=0$ LL is one-half filled and the 2DES is well-approximated as a Fermi liquid of composite fermions in zero effective (orbital) magnetic field.  As reported previously \cite{ying,chickering}, at sufficiently low temperatures \Sxx\ depends linearly on the temperature $T$.  In this regime phonon drag effects are negligible and the thermopower is dominated by diffusion within the 2DES.  The situation resembles zero magnetic field where the standard Cutler-Mott theory of thermopower applies \cite{mott}.  Figure 1 shows our present high-precision \Sxx\ data at $B = 8.1$ T where $\nu =3/2$.  A clear cross-over from a linear to super-linear temperature dependence is seen.  In the super-linear regime above $T \sim 200$ mK phonon drag effects are likely becoming significant.  However, the low Fermi temperature $T_F$ of the composite fermion metal at this filling suggests that finite temperature corrections to the standard Cutler-Mott result for the diffusion thermopower could also be involved.  Following Cooper, Halperin, and Ruzin \cite{cooper}, the linear temperature dependence of \Sxx\ in the diffusive regime can be used to estimate the mass $m_{CF}$ of the composite fermions at $\nu = 3/2$; we find $m_{CF} \approx 1.7 m_e$, where $m_e$ is the bare electron mass \cite{cfmass}.
\begin{figure}
\begin{center}
\includegraphics[width=1.0 \columnwidth] {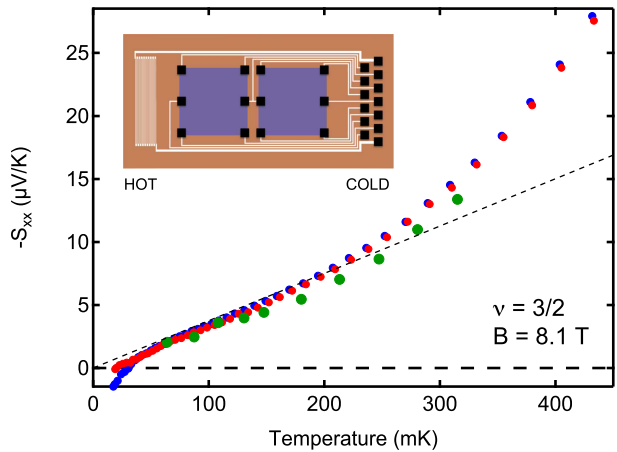}
\end{center}
\caption{(color online). Thermopower vs. temperature at $\nu = 3/2$.  Red and blue dots correspond to data from the two 2DES mesas on the sample bar.  Green dots are from a previous measurement on a different sample \cite{chickering}. 
Inset shows device layout, with serpentine heater at the left and contact patch bay at right.}
\label{fig1}
\end{figure}

For temperatures above about $T=40$ mK the thermopower data extracted from the two independent 2DES mesas are virtually identical.  Below this temperature the two data sets differ and individually begin to deviate from simple $T$-linear behavior.  Indeed, the observed thermovoltages can even change sign at very low $T$.  We attribute this behavior to effects analogous to universal conductance fluctuations (UCFs) in disordered mesoscopic conductors \cite{lee,esposito,gallagher}.  In this low temperature regime we find that the thermoelectric voltages can fluctuate aperiodically with magnetic field in the field regime around $\nu = 3/2$ and display strong sensitivity to the thermal history of the sample. Both of these signatures are familiar from UCF phenomenology.  Above 40 mK these mesoscopic thermopower anomalies rapidly subside.   

Figure 2a shows the measured longitudinal thermopower \Sxx\  between filling factors $\nu = 2$ and $\nu = 3$; i.e. in the lower spin branch of the $N=1$ Landau level, at $T=200$ mK and $T = 60$ mK.  These data vividly reveal thermoelectric signatures of fractional quantized Hall states at $\nu = 5/2$, 7/3, 8/3, 11/5, and 14/5, with deep minima in \Sxx\ at the appropriate magnetic fields developing as the temperature is reduced.  This is, of course, the expected behavior for gapped quantized Hall states.  We emphasize that the broad zeroes of \Sxx\ in the integer quantized Hall states at $\nu = 2$ and 3 are genuine; no spurious signals have been subtracted from these data.  This demonstrates that extrinsic thermoelectric signals in our measurement circuit are negligible.  We find \Sxx\ to be remarkably quantitatively consistent from one cool-down of the sample to the next and to be essentially independent of which of the three possible contact pairs on each mesa is used to detect the thermovoltage. These observations contrast with ordinary electrical transport which often shows significant sensitivity to thermal cycling and choice of contacts.  We speculate that this consistency reflects the connection of \Sxx\ to the thermodynamic entropy of the 2DES rather than the details of charge transport through the system.

The temperature dependence of \Sxx\ at $\nu = 5/2$ and 7/3 is shown in Fig. 2b.  From about $T = 200$ mK down to roughly $T = 50$ mK \Sxx\ is thermally activated ($S_{xx} \sim e^{-\Delta/2T}$) at both of these fillings.  Energy gaps of $\Delta \approx 430$ mK, for both $\nu = 5/2$ and 7/3,  are estimated from the data.  Statistically identical values are obtained from the two 2DES mesas.  These gap values are consistent with prior determinations \cite{rxxgap} based on measurements of the longitudinal resistivity, $\rho_{xx}$.  At lower temperature \Sxx\ deviates from simple thermally activated behavior, suggesting the increasing importance of hopping between localized electronic states.  Similar low temperature deviations from thermal activation are commonplace in resistivity measurements in quantized Hall states.
\begin{figure}
\begin{center}
\includegraphics[width=1.0 \columnwidth] {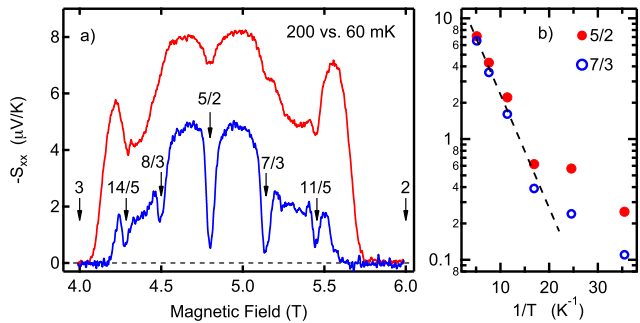}
\end{center}
\caption{(color online). a) \Sxx\ between $\nu = 2$ and $\nu = 3$ at $T = 200$ mK (red) and $T = 60$ mK (blue) with various quantized Hall states indicated by arrows.  b) \Sxx\ at $\nu = 5/2$ and $\nu = 7/3$ vs. $1/T$.  Dashed line corresponds to an energy gap of $\Delta = 430$ mK.}
\label{fig2}
\end{figure}

As the temperature is reduced below about $T = 40$ mK substantial changes are observed in the thermopower in the $N= 1$ Landau level.  These changes are most dramatic at magnetic field locations between the prominent fractional quantized Hall effects at $\nu = 5/2$ and $\nu = 7/3$ and between $\nu = 5/2$ and $\nu = 8/3$.  Figure 3a illustrates the change in \Sxx\ which develops between $T = 41$ mK and $T = 28$ mK.  In particular around $B = 4.67$ T and $B = 4.93$ T deep minima appear in \Sxx.  These minima develop quite suddenly as functions of temperature, with Fig. 3b showing the temperature dependence of \Sxx\ at $B = 4.67$ T.   Interestingly, as Fig. 3b shows, at temperatures just above the transition, the thermopower is roughly temperature independent.  This is a strong deviation from the linear temperature dependence of \Sxx\ in ordinary metals and in the compressible metallic phases of the 2DES at $\nu = 1/2$ and $\nu = 3/2$ in the lowest Landau level \cite{ying,chickering,cooper}.

\begin{figure}
\begin{center}
\includegraphics[width=1.0 \columnwidth] {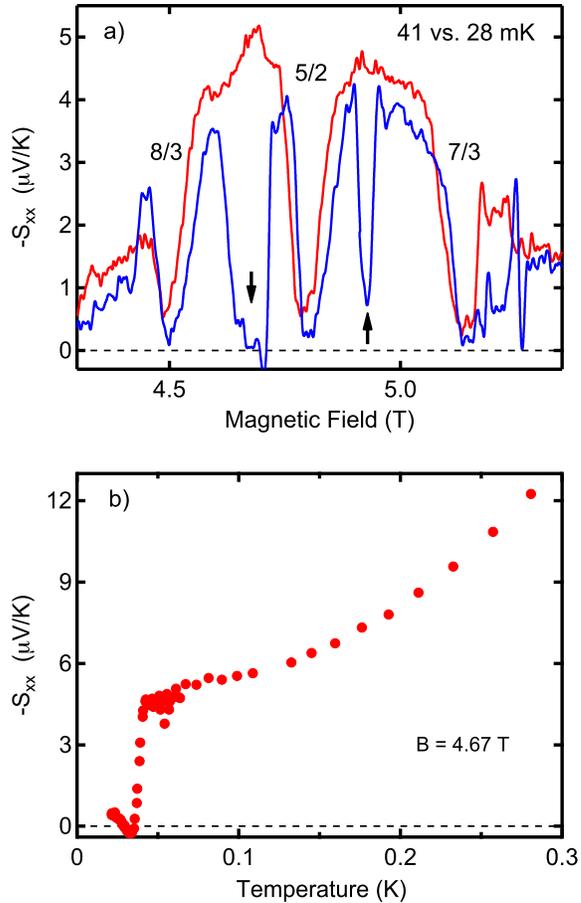}
\end{center}
\caption{(color online). a) \Sxx\ between $\nu = 2$ and $\nu = 3$ at $T = 41$ mK (red) and $T = 28$ mK (blue). Arrows indicate field locations where a particularly strong temperature dependence is observed.  b) \Sxx\ vs. $T$ at $B = 4.67$ T.}
\label{fig3}
\end{figure}
We believe that these new low temperature minima in \Sxx\ are associated with the re-entrant integer quantized Hall effect (RIQHE) phases known to exist in the $N = 1$ Landau level \cite{riqhe}.  These states are insulating configurations of the 2D electrons in the $N=1$ Landau level that appear between the major incompressible FQHE liquid phases at $\nu = 5/2$, 7/3, 8/3, 11/5 and 14/5.  Qualitatively similar to the so-called ``bubble'' phases which exist in the $N \geq 2$ Landau levels \cite{kfs,mc}, the RIQHE phases in the $N=1$ Landau level likely possess at least short-range crystalline order.  Recent transport experiments have revealed that the collapse of the resistivity $\rho_{xx}$ to zero and the formation of the integer quantized Hall plateau is remarkably abrupt as a function of temperature \cite{csathy}.  We find that these resistive transitions coincide with the collapse of \Sxx\ that we are reporting here.  

Similarly abrupt collapses of \Sxx\ are seen at other RIQHE states in the $N=1$ Landau level, including those adjacent to the $\nu = 7/2$ FQHE state in the upper spin branch of the Landau level.  In some cases \Sxx\ first shows a local maximum before falling toward zero as $T\rightarrow 0$.  These thermopower results, coupled with recent resistivity measurements \cite{csathy} and prior reports of magnetic field hysteresis and non-linear transport signatures \cite{riqhe}, strongly suggest that the RIQHE develops via finite-temperature first-order phase transitions within the 2DES.   

We note in passing that at the lowest temperatures ($T \sim 20$ mK) \Sxx\ can change sign and become positive in narrow ranges of magnetic field (the conventional sign of \Sxx\ for electrons is negative).  These sign changes are most obvious in the $N =1$ Landau level, but are evident elsewhere as well (as Fig. 1 demonstrates at $\nu = 3/2$).   Such sign changes are not surprising since \Sxx\ depends essentially on the $derivative$ $\partial \sigma / \partial \mu$ of the conductivity with respect to chemical potential \cite{mott,girvin}.  In a mesoscopic sample universal conductance fluctuations produce a distinct ``magneto-fingerprint'' and thus sign changes in \Sxx\ versus magnetic field are certainly expected.  In the RIQHE in the $N=1$ Landau level, the dramatic non-monotonicity of the Hall resistance $\rho_{xy}$ (and hence Hall conductivity $\sigma_{xy}$) might also be expected to produce sign changes in \Sxx\ via the generalized Mott formula first advanced by Jonson and Girvin \cite{girvin}. 

\begin{figure}
\begin{center}
\includegraphics[width=0.9 \columnwidth] {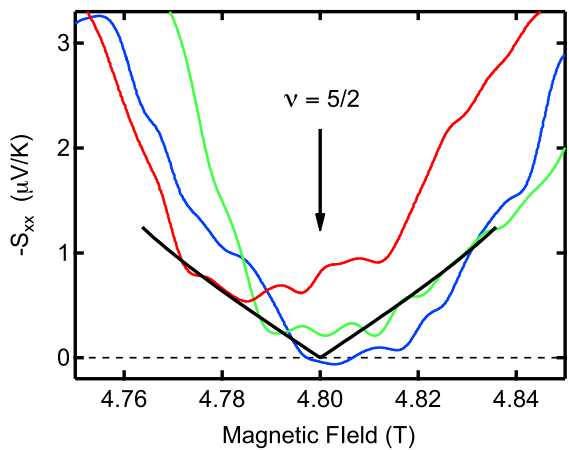}
\end{center}
\caption{(color online). \Sxx\ near $\nu = 5/2$ at $T = 20$ mK (blue), 28 mK (green) and 41 mK (red).  Solid black lines give theoretical prediction of Yang and Halperin \cite{yang}.}
\label{fig4}
\end{figure}
Finally, we return to $\nu = 5/2$ where recent theoretical work \cite{yang} has suggested that \Sxx\ might reflect the anticipated non-abelian exchange statistics of the quasiparticles excitations of the Moore-Read ground state \cite{read}.
Yang and Halperin \cite{yang} have predicted that owing to the ground state degeneracy of an ensemble of non-abelian quasiparticles, the thermopower \Sxx\ near $\nu = 5/2$ will, in the clean limit, be temperature independent and proportional to $|B-B_0|$, the deviation of the magnetic field from its value $B_0$ at $\nu= 5/2$.  Quantitatively, they find $S_{xx} = - |(B-B_0)/B_0|(k_B/|e^*|)$log $d$, where $k_B$ is Boltzmann's constant, $e^*=e/4$ the quasiparticle charge, and $d = \sqrt2$ the quantum dimension appropriate to the Moore-Read state (or its particle-hole conjugate).  This result is expected to be valid at temperatures low enough that other sources of entropy are frozen out but also high enough to smear the tunnel splittings between quasiparticles that will ultimately lift the ground state degeneracy.  

Figure 4 compares our low temperature \Sxx\ data near $\nu = 5/2$ with the theoretical prediction of Yang and Halperin \cite{yang}.  While our data are quantitatively roughly consistent with the theory, they do not offer compelling support for it.  At these low temperatures, the sub-$\mu$V/K thermopower results in extremely small thermoelectric voltages ($\lesssim 2$ nV at $T = 20$ mK with $\Delta T = 2$ mK).  Substantial signal averaging is thus required and renders the data sensitive to long-term drifts in the measurement set up.  Indeed, as Fig. 4 shows, the location of the $\nu = 5/2$ minimum is not precisely the same at each temperature; small, history-dependent shifts of FQHE minima were frequently encountered.   These difficulties prevented us from cleanly observing the expected temperature independence of \Sxx\ on the flanks of the $\nu = 5/2$ minimum.  Future experiments, possibly employing the recently suggested Corbino geometry \cite{barlas}, may improve the situation.

In conclusion, we have presented measurements of the diffusion thermopower of an ultra-high mobility two-dimensional electron system in the fractional quantum Hall regime at temperatures down to $T = 20$ mK.  Our data clearly reveal thermoelectric signatures of several fractional quantized Hall states in the $N = 1$ Landau level.  At $\nu = 5/2$ and $\nu = 7/3$ the temperature dependence of the thermopower implies energy gaps of approximately $\Delta \approx 430$ mK for both states.  More strikingly, we observe abrupt collapses of the thermopower coincident with the appearance the insulating RIQHE phases in both spin branches of the $N = 1$ Landau level.

This work was supported via DOE grant DE-FG03-99ER45766 and Microsoft Project Q.


\begin{references}
\bibitem{geometry} The finite sizes of both the 2DES mesas and the indium solder joint to the cold finger do result in some systematic uncertainty in the both the deduced thermal conductances and thermopowers.   We estimate these geometric uncertainties to be around 10$\%$.
\bibitem{timeconstant} The long time constants encountered previously \cite{chickering} were due to nuclear heat capacities in the manganin sample wiring and indium dots used for ohmic contacts.  The present all-evaporated design vastly reduces this excess heat capacity.
\bibitem{cfmass} As discussed previously \cite{cooper,chickering}, this estimate assumes complete spin polarization of the 2DES at $\nu =3/2$.  If the 2DES is completely unpolarized, the mass becomes $m_{CF} = 0.85 m_e$.
\end{references}
\end{document}